# Integration and commissioning plan of Full Flow Purifier at Muon Campus

**J. Subedi, T. Tope, B. Hansen, M. White, J. Tillman, V. Patel, W. Cyko, J. Dong**

Fermi National Accelerator Laboratory, Batavia, IL, USA

Email: jsubedi@fnal.gov

**Abstract**. The Full Flow Purifier for Fermilab's Muon Campus uses a charcoal bed surrounded by a liquid nitrogen jacket to purify up to 240 g/s of helium gas. Fabrication by Ability Engineering Technology Inc. has been completed and the purifier delivered to Fermilab. It is the largest purifier to be used at Fermilab based on both capacity and size. A previous paper discussed the design of purifier for various operational conditions and horizontal shipping. The purifier is designed to withstand 5g force in vertical and 2g force in lateral and longitudinal directions. Transportation experience from vendor to Fermilab and within site is discussed. Integration of the purifier involved design and fabrication of a liquid nitrogen transfer line, regeneration system, and helium piping to connect it to Muon Campus cryogenic system. It also involved establishing electrical, instrumentation and controls connections to the system. Integration of the purifier is discussed in detail. The purifier is to be commissioned using up to 4 MYCOM helium compressors to supply helium and liquid nitrogen supplied by a 60,000-liter tank. Actual effectiveness of the 3-stream heat exchanger is to be estimated based on measured temperatures and flow rate. Impurity levels will be monitored at inlet and outlet of the purifier. Theoretical adsorption capacity of the purifier is calculated based on the measured temperatures and flowrates and is compared to actual adsorption capacity over time.

## 1. Background

The experiments at Muon Campus require stable helium supply at 5 K to the superconducting magnets. Previously after a short period of operation at Muon Campus, the pressure-drop across the magnet's flow supply valve increased frequently due to impurities, resulting in reduced excess capacity concurrently with drop in refrigerator expander efficiency. This required periodic powering down of the magnets to allow for a helium valve "flush" to remove accumulated contamination at the valve as well as warming up the refrigerator expanders above 80 K to release contamination. A "mobile" purifier was used for this periodic process but only at 10% of total refrigerator flow such that the purification process is extended. After long duration of running, this purification procedure was required almost every two weeks to restore proper liquid helium supply to the magnet. This led to a requirement for a full flow purifier which can purify impurities from entire Muon Campus refrigerator system and mitigate existing impurities issues and associated experiment downtime for the future Mu2e experiment. A previous paper [1] discussed on the design, sizing and FEA of the purifier for operational and shipping conditions. Fabrication of this purifier was completed by Ability Engineering Technology Inc. and was delivered to Fermilab in November 2024. This paper focuses on shipping and lifting experience of the purifier, acceptance testing, integration effort and commissioning plans of the purifier.

## 2. Transportation of purifier

During the design phase, FEA analysis of purifier was performed to ensure it withstands shipping loads during horizontal shipping. Shipping pads were added to restrict movement of internal components. Heat exchanger rotational stops and shipping restraints were also added to ensure internal components can withstand 2g lateral and longitudinal load and 5g vertical load during its shipment.

The purifier was fabricated by Ability Engineering Technology Inc. in the horizontal orientation. For loading the purifier for transportation from the vendor to Fermilab, two fork trucks were used by vendor to lift the purifier. The purifier was lifted along with the wooden cribs tied to vacuum shell as shown in **Figure 1** which made it easier to lower the purifier to the trailer bed. Transportation was done using Conestoga flatbed trailer with air ride suspension. Two Shocklog ® 298 Impact recorder and datalogger shockloggers were installed at both ends of the purifier to monitor shock loading during lifting and transportation. The purifier was transported at an average speed of 45 mph from the vendor to Fermilab. Shockloggers were removed and the loading data during transportation were analyzed which showed maximum load in any direction during shipping and loading to be less than 1 g.

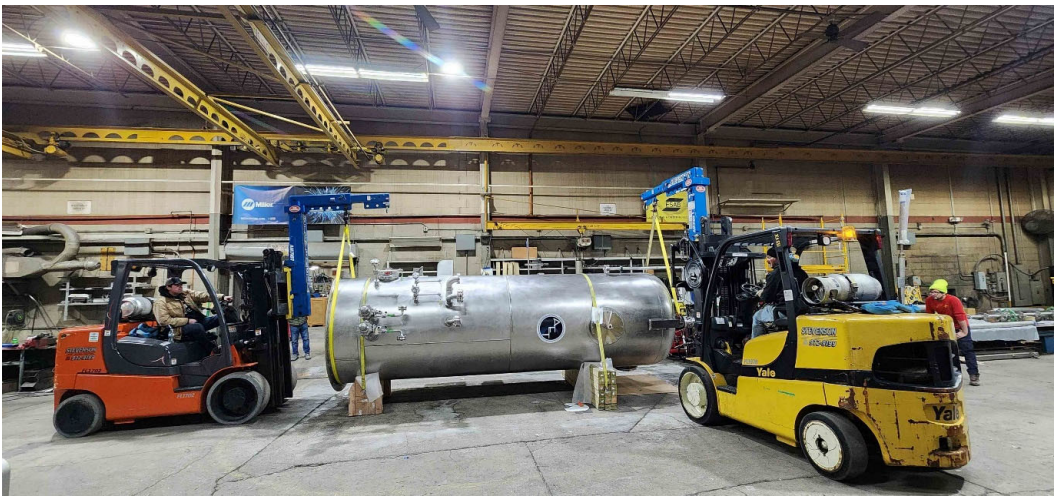

**Figure 1.** Lifting of purifier along with wooden cribs using fork trucks at Ability Engineering Technology Inc. before it is loaded to the trailer for transport.

The purifier was delivered to an indoor testing location at Fermilab where acceptance testing was performed. Post acceptance test, the purifier had to be transported a distance of around 2 miles onsite to its final installation location. Due to use of brittle material (e.g. G10) for shipping support and due to the cost of replacement, the purifier onsite transportation was designated as critical transport. All documentation required for critical transportation of the purifier including transportation specification, FMEA, and checklist was provided to transportation review panel. Two independent FEA analysis showing ability of purifier to handle 5g vertical and 2g lateral and longitudinal loads along with shocklogger report for transportation from vendor to Fermilab were also presented to the panel which showed maximum shock during lifting and transport was less than 1g in any direction. These documentations were reviewed by transportation review panel and critical transportation of the purifier was recommended. After critical transportation approval, purifier was transported from testing location to installation location following critical transportation plan.

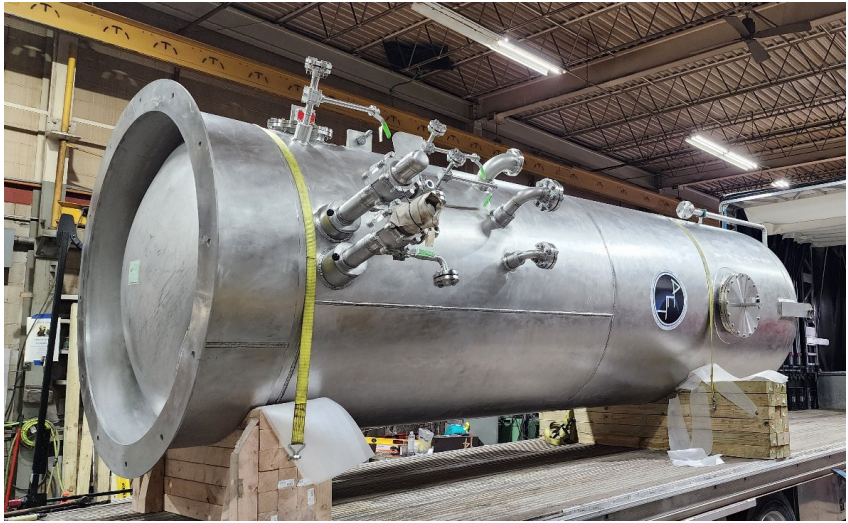

**Figure 2.** Horizontal shipping of the purifier from vendor to Fermilab in wooden cribbing with shockloggers attached on two ends

## 3. Acceptance Testing of Purifier

After purifier was transported from the vendor to Fermilab, it was imperative to perform acceptance testing to ensure structural integrity and leak tightness of purifier components after shipping. After the purifier was delivered to an indoor testing location at the D0 building at Fermilab, instrumentation gauges were connected to the circuits and insulating vacuum space was pumped down on the purifier. The rate of vacuum rise on insulating vacuum was performed on purifier overnight by isolating the vacuum pump after vacuum readback was stabilized. The insulating vacuum pressure remained stable overnight after few pump-down and backfill cycles to remove moisture.

The specification for acceptance leak test was that any increase in leak rate above $1 \times 10^{-9}$ torr*l/s from background would be determined as leak. Once the insulating vacuum pressure stabilized, insulating vacuum space was connected to the mass spectrometer leak detector (MSLD). Helium was sprayed on all the outdoor welds of vacuum vessel. Purifier vacuum vessel was bagged using a plastic envelope and helium was sprayed on the envelope. No increase in leak rate from background was seen on purifier.

After the vacuum circuit leak test was successfully completed, leak test of helium and nitrogen circuit at Maximum Allowable Working Pressure (MAWP) was performed. For this test, insulating vacuum remained pumped down and connected to the MSLD and a background reading was taken. Once background reading was stable and below $10^{-9}$ torr*l/s, helium-nitrogen mixture was introduced into each circuit separately and pressurized to MAWP. The leak detector reading is taken 5 minutes after MAWP of the circuit is reached. If the leak rate increased above $1*10^{-9}$ torr*l/s, it would be deemed as a leak and rectification measures would be taken. **Table 1** shows the leak rate measured after introduction of helium at MAWP in the circuit. The leak test results showed no increase in leak rate from background after introduction of helium in circuits. This satisfied the criteria for acceptance test after which purifier was transported to its installation location for vertical installation.

**Table 1.** Acceptance leak test result for different circuits of the purifier.

| Circuit | Circuit pressure | Surrounding pressure | Measured leak rate |
| --- | --- | --- | --- |
| Insulating vacuum space | 15 mTorr | 1 atm | 5.5 x$10^{-10}$ torr*l/s |
| Helium circuit | 350 psig | 15 mTorr | 8.0 x$10^{-10}$ torr*l/s |
| Nitrogen circuit | 150 psig | 15 mTorr | 3.8 x$10^{-10}$ torr*l/s |

## 4. Lifting of Purifier

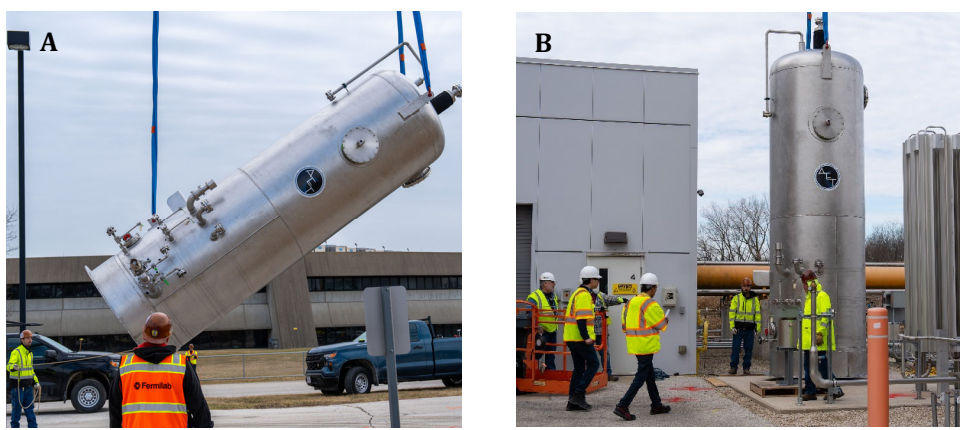

**Figure 3.** (A) Lifting of purifier using two cranes for installation and (B) final installation of purifier in vertical orientation (Photo credit: Caitlyn Buongiorno)

Design of lifting lugs was conducted such that they are rated for weight distribution of purifier in vertical, horizontal or any angular orientation. Lifting lugs analysis was performed by Ability Engineering Technology Inc. and reviewed by Fermilab to ensure appropriate load distribution and considerations are taken for lifting. Due to shipping of the purifier in a horizontal orientation and need to install it in vertical orientation for operation, two cranes were used to lift purifier horizontally on air after untying it from wooden cribbing. Orientation of purifier was gradually changed from horizontal to vertical by lowering one crane as shown in figure 3(A). Finally, purifier was oriented vertically and two lugs on top of the purifier shared the entire load. Sling and shackles connected to bottom lugs were disconnected once purifier reached vertical orientation. Use of two cranes mandated the lifting to be a critical lift. For critical lifts, proof tested shackles and slings were used.

## 5. Integration of purifier

After installation of the full flow purifier, several efforts are underway to integrate it to Muon Campus cryogenic system. Helium piping is installed to connect the purifier to common discharge of 4 MYCOM compressors. A final filter, which filters fine charcoal particles coming out of purifier, is installed on the outlet of the purifier. The purifier and associated internal piping were tested at

121 % of MAWP. Both rupture disks and relief valves are installed in parallel for each pressure vessels such that rupture disk act as primary relief device and has a burst pressure higher than the relief valve, which is present just to avoid nuisance and inventory losses associated with blowing of rupture disk. This waives relief testing requirement every 5 years. Integration of liquid nitrogen piping is done in such a manner that liquid nitrogen can be supplied to the purifier using one of the two existing and future 60,000-liter liquid nitrogen tanks. The liquid nitrogen transfer line also has provision to connect to a mobile purifier which can be used for helium purification during regeneration of full flow purifier. Integration of regeneration piping is underway which connects gaseous nitrogen from tank to the regeneration port of purifier. Regeneration heater is integrated to warm up nitrogen gas before being sent to the purifier.

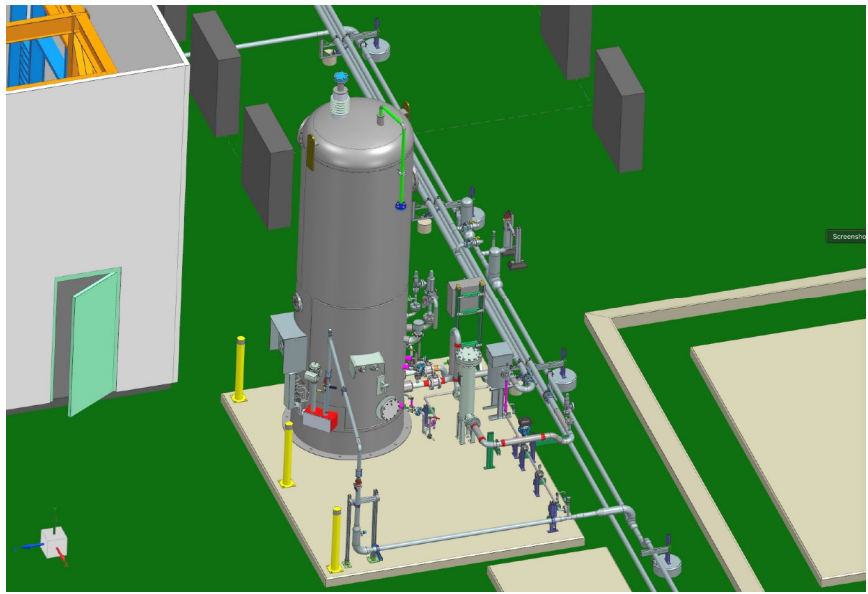

**Figure 4.** Integration model for purifier showing piping and components.

Appropriate control features like liquid level control, regeneration flow control, regeneration temperature control are being incorporated in the system. Several instrumentation and controls integration for purifier is underway. Outdoor rated transmitters and electrical components are installed for process of monitoring and control. All readbacks and control loops are integrated to iFix such that operations off purifier can be controlled remotely. Control loops and trip functions are enabled such that the purifier remains in safe mode during operations.

## 6. Commissioning plan

After integration of full flow purifier is completed, commissioning of purifier shall take place in Fall 2025. Procedures are prepared for cooldown and regeneration commissioning of the purifier. Before operations, activated charcoal on the purifier needs to be regenerated due to ingression of air and moisture. Thus, regeneration commissioning shall be performed prior to cooldown by supplying dry nitrogen on helium adsorber vessel through a regeneration heater. Once moisture is removed which is confirmed by checking dew point level of outgoing gas, the system is pumped and backfilled 3 times between pressures of 10 torr and 1 psig using helium inventory.

Cooldown of the purifier is performed gradually for commissioning. Helium is pressurized to operating pressure on adsorber vessel. Liquid nitrogen control valve is gradually opened to cool down the nitrogen jacket and then fill the jacket with liquid nitrogen. The liquid level will be maintained at 80 % of total jacket level by using differential pressure as process variable for control loop on LN2 supply valve. 4 MYCOM compressors shall be used for commissioning of the purifier. Helium flow shall be established gradually by closing on the bypass valve until all helium inventory flow is circulated through the purifier. The 3-stream heat exchanger shall gradually cooldown which can be monitored by the temperature on both warm end and cold end. Arc cell spectrographic detector which measures nitrogen contamination level shall be used on both inlet and outlet stream of the purifier. Actual adsorption capacity of the purifier can be calculated over time based on the flowrate and inlet and outlet impurities and compared with the theoretical capacity. Based on the measured flowrate and temperatures, effectiveness of heat exchanger shall be calculated.

## 7. Summary

The design aspect of the purifier was described in detail in the previous paper [1]. Transportation of the purifier, acceptance testing criteria, procedure and results are discussed. Approval process and steps for shipping and lifting of purifier onsite along with final installation are discussed. Integration work is being carried out post installation of purifier at its final location and currently commissioning plan is being prepared.

## 8. References

[1]   Subedi, Jeewan, et al. "Design of Muon Campus full flow purifier for varying operational conditions and horizontal shipping." *IOP Conference Series: Materials Science and Engineering*. Vol. 1301. No. 1. IOP Publishing, 2024.

**Acknowledgments**